\documentclass[aps,letterpaper,prl,twocolumn,showpacs,nofootinbib]{revtex4-1}
\usepackage{amssymb}
\usepackage{amsmath,bm}
\usepackage[pdftex]{graphicx,color}
\usepackage{hyperref}
\usepackage{enumitem}
\usepackage{epstopdf}
\begin{document}

\author{Zongda Li$^{1,2}$}
\email{zongda.li@auckland.ac.nz}
\author{Yiqing Xu$^{1,2}$}
\author{Caleb Todd$^{1,2}$}
\author{Gang Xu$^{1,2}$}
\author{St\'ephane Coen$^{1,2}$}
\author{Stuart G. Murdoch$^{1,2}$}
\author{Miro Erkintalo$^{1,2}$}
\affiliation{$^1$Department of Physics, University of Auckland, Auckland 1010, New Zealand}
\affiliation{$^2$The Dodd-Walls Centre for Photonic and Quantum Technologies, New Zealand}

\title{Observations of Existence and Instability Dynamics of Near-Zero-Dispersion Temporal Kerr Cavity Solitons}

\begin{abstract}

\noindent Dissipative Kerr cavity solitons (CSs) are persisting pulses of light that manifest themselves in driven optical resonators and that have attracted significant attention over the last decade. Whilst the vast majority of studies have revolved around conditions where the resonator exhibits strong anomalous dispersion, recent studies have shown that solitons with unique characteristics and dynamics can arise under conditions of near-zero-dispersion driving. Here we report on experimental studies of the existence and stability dynamics of Kerr CSs under such conditions. In particular, we experimentally probe the solitons' range of existence and examine how their breathing instabilities are modified when group-velocity dispersion is close to zero, such that higher-order dispersion terms play a significant role. On the one hand, our experiments directly confirm earlier theoretical works that predict (i) breathing near-zero-dispersion solitons to emit polychromatic dispersive radiation, and (ii) that higher-order dispersion can extend the range over which the solitons are stable. On the other hand, our experiments also reveal a novel cross-over scenario, whereby the influence of higher-order dispersion changes from stabilising to destabilising. Our comprehensive experiments sample soliton dynamics both in the normal and anomalous dispersion regimes, and our results are in good agreement with numerical simulations and theoretical predictions.

\end{abstract}
\maketitle

\section{Introduction}

Temporal Kerr cavity solitons (CSs) are localized pulses of light that can circulate in a driven passive optical resonator without distortion~\cite{wabnitz_suppression_1993, leo_temporal_2010}. They underpin the highly-coherent Kerr frequency combs that can be generated in low-loss microresonators~\cite{coen_modeling_2013, erkintalo_coherence_2014,herr_temporal_2014, kippenberg_dissipative_2018, pasquazi_micro-combs_2018} and have enabled ground-breaking advances in a wide array of applications, including telecommunication~\cite{marin-palomo_microresonator-based_2017}, frequency synthesis~\cite{spencer_optical-frequency_2018}, optical ranging~\cite{trocha_ultrafast_2018} and many more~\cite{suh_microresonator_2016,hu_reconfigurable_2020,liang_high_2015,papp_microresonator_2014}. CSs are also dynamically rich. In particular, for certain system parameters, CSs are known to destabilise through a Hopf bifurcation, giving rise to oscillatory behaviour~\cite{leo_dynamics_2013, lucas_breathing_2017, yu_breather_2017, yi_imaging_2018}. Also more complex instability behaviours~\cite{johansson_stability_2019,xu_spontaneous_2020,yang_stokes_2017,jang_controlled_2016}, including spatiotemporal chaos~\cite{anderson_observations_2016}, soliton binding~\cite{wang_universal_2017, weng_heteronuclear_2020} and soliton crystals~\cite{cole_soliton_2017, guo_universal_2017, karpov_dynamics_2019}, are possible.

Most of the studies concerning CSs have, to date, revolved around the scenario where the cavity driving field experiences strongly anomalous group-velocity dispersion, with higher-order dispersion terms acting as small perturbations only. In this regime, the most conspicuous perturbative effect of higher-order dispersion is the generation of dispersive radiation that manifests itself as a single narrow-band Lorentzian peak in the spectral domain~\cite{jang_observation_2014, milian_soliton_2014, brasch_photonic_2016, cherenkov_dissipative_2017}. More recently, there has however been growing interest in exploring the dynamics of CSs under conditions of near-zero-dispersion driving, i.e., with the wavelength of the input laser coincident (or close to coincident) with the zero-dispersion point of the resonator. In this regime, higher-order dispersion dominates the dynamics, giving rise to novel bright localized structures both in the normal and anomalous dispersion regimes~\cite{parra-rivas_coexistence_2017,li_experimental_2020,anderson_zero-dispersion_2020}. Theoretical studies have also suggested that near-zero-dispersion conditions can significantly affect the solitons' instability dynamics and range of existence~\cite{melchert_multi-frequency_2020,parra-rivas_third-order_2014}. In particular, it has been theoretically predicted that (i) strong third-order dispersion combined with soliton breathing can lead to the emission of comb-like, \emph{polychromatic} dispersive radiation~\cite{melchert_multi-frequency_2020} -- in stark contrast to the sharp, \emph{monochromatic} dispersive radiation that is commonly observed with stable (i.e. non-oscillatory) solitons~\cite{jang_observation_2014,li_experimental_2020} -- and that (ii) third-order dispersion can stabilize CSs, i.e., restrict the range of  parameters over which the solitons are oscillatorily  unstable~\cite{parra-rivas_third-order_2014}. However, as of yet, no comprehensive experimental tests of these predictions have been reported.

In this Article, we experimentally explore the existence and stability dynamics of Kerr CSs under conditions of near-zero-dispersion driving. Our experiments are performed in a coherently driven fibre ring resonator, where the dispersion conditions are systematically controlled by tuning the driving wavelength~\cite{li_experimental_2020}. Our results provide experimental confirmation of the theoretical predictions mentioned above~\cite{melchert_multi-frequency_2020, parra-rivas_third-order_2014}. First, we observe the emission of polychromatic dispersive radiation emitted by breathing solitons. Second, we confirm that third-order dispersion indeed can stabilise CSs~\cite{parra-rivas_third-order_2014}. Interestingly, however, our experiments also reveal for the first time that the latter behaviour is not universal: the impact of higher-order dispersion switches from stabilising to destabilising for high driving power levels. Our results provide significant new insights into the dynamics of CSs under conditions of near-zero-dispersion driving and could facilitate the design of broadband frequency combs or novel sources of ultra-short pulse trains.

\section{Experimental and numerical methods}

\begin{figure}[!h]
	\centering
	\includegraphics[scale=1]{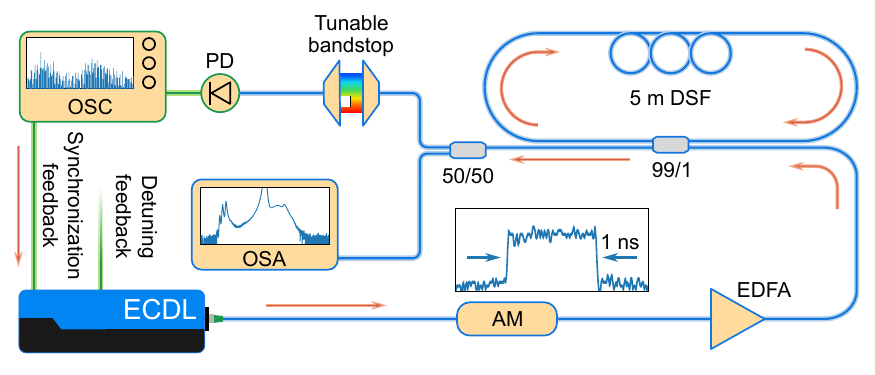}
	\caption{\small Simplified depiction of the experimental setup. AM, amplitude modulator; DSF, dispersion-shifted fiber; ECDL, external-cavity diode laser; EDFA, Erbium-doped fiber amplifier; OSA, optical spectrum analyzer; OSC, oscilloscope. }\label{schematic}
\end{figure}

Figure~\ref{schematic} shows a schematic illustration of our experimental setup, which is overall similar to that used in Ref.~\cite{li_experimental_2020}. The experiment is built around a 5-m-long optical fibre ring resonator with a free-spectral range of 41.5~MHz and Kerr nonlinearity coefficient \mbox{$\gamma = 1.8~\mathrm{W}^{-1}\mathrm{km}^{-1}$}. The zero-dispersion wavelength (ZDW) of the resonator was measured to be 1564.5~nm with third- and fourth-order dispersion coefficients \mbox{$\beta_{3,\mathrm{zdw}} = 0.13~\mathrm{ps}^3\mathrm{km}^{-1}$} and \mbox{$\beta_{4,\mathrm{zdw}} = -6.65\times10^{-4}~\mathrm{ps}^4\mathrm{km}^{-1}$} at the ZDW, respectively.  The cavity is comprised of a single 99/1 fibre coupler spliced to form a ring, yielding a large measured finesse of \mbox{$\mathcal{F} = 400$}. We synchronously drive the resonator with a train of nanosecond pulses carved from a tunable external cavity diode laser (ECDL). As described in Ref.~\cite{li_experimental_2020}, the synchronization between the nanosecond driving pulses and the sub-picosecond solitons inside the resonator is actively maintained by using a computer-controlled feedback loop. The driving pulses are amplified and injected into the resonator via the input port of the 99/1 coupler and the intracavity dynamics are monitored both in the temporal and spectral domains at the output port of that coupler. Because the output consists of a superposition between the intracavity field and the reflected part (99\%) of the driving field, a tunable bandstop filter is used to remove the latter prior to temporal measurements so as to improve the signal-to-noise ratio. To overcome fluctuations in the frequencies of the driving laser and the cavity modes, we actively lock the driving laser to a cavity resonance by using the schemes introduced in~\cite{nielsen_invited_2018} and \cite{li_experimental_2020}. This allows us to systematically control the detuning between the laser and the cavity, and hence investigate the detuning-dependent properties of the CSs.

To compare our experimental observations with theoretical predictions, we model our system using the generalized Lugiato-Lefever equation (LLE) that includes higher-order dispersion and delayed Raman nonlinearity~\cite{wang_stimulated_2018}:
	\begin{equation}
		\small
		\begin{split}
		t_R\frac{\partial E(t,\tau)}{\partial t}&=\left[ -\alpha-i\delta_\mathrm{0}+\sum_{k=2}^4\frac{i^{k+1}}{k!}L\beta_k\frac{\partial^k}{\partial \tau^k}\right]E+\sqrt{\theta}E_{\text{in}}\\&+i\gamma L\left[(1-f_R)|E|^2+f_Rh_R(\tau)*|E|^2\right]E.
		\end{split}
		\label{LLE}
	\end{equation}
Here, $t_\mathrm{R}$ is the round trip time, $E(t,\tau)$ is the slowly varying intracavity electric field envelope with units of $W^{1/2}$, $t$ is a slow time that describes the evolution of $E$ at the scale of the photon lifetime, $\tau$ is a corresponding fast-time which describes the intracavity field temporal profile over a single roundtrip, $\alpha = \pi/\mathcal{F}$ is half of the fraction of power lost per round trip, $\delta_0$ is the phase detuning of the driving field from the closest cavity resonance, $\beta_k$ are the dispersion coefficients at the pump frequency, $L$ is the cavity length, $\theta$ is the power transmission coefficient of the input coupler ($\theta = 0.01$ in our experiments), $E_\mathrm{in} = \sqrt{P_\mathrm{in}}$ is the amplitude of the driving field with power $P_\mathrm{in}$, $f_\mathrm{R} = 0.18$ is the Raman fraction of the nonlinearity, $h_\mathrm{R}(\tau)$ is the corresponding time-domain Raman response function~\cite{hollenbeck_multiple-vibrational-mode_2002} and $*$ denotes convolution. Note that our model includes fourth-order dispersion and Raman nonlinearity to improve quantitative agreement with our experiments, but we have carefully verified that they do not affect the overall trends that are instead set by the third-order dispersion $\beta_3$. Furthermore, in what follows,  we will discuss and contrast the results obtained from Eq.~\ref{LLE} with corresponding results of the `pure' LLE that is obtained from Eq.~\ref{LLE} by setting all higher-order terms to zero, i.e., $\beta_k = 0$ for $k>2$ and $f_\mathrm{R} = 0$. 

In our analysis below, we will refer to the following normalized parameters~\cite{leo_temporal_2010, jang_observation_2014}: driving power \mbox{$X = \gamma P_\mathrm{in}L\theta/\alpha^3$}, detuning $\Delta = \delta_0/\alpha$, and third-order dispersion \mbox{$d_3 = \sqrt{2\alpha/L}\beta_3/(3|\beta_2|^{3/2})$}. We stress that the normalized parameter $d_3$ describes the relative strength of third-order dispersion; for $|d_3| \geq 1$, the impact of third-order dispersion is comparable or dominates over that of second-order dispersion~\cite{jang_observation_2014}. Thanks to the use of a widely tunable ECDL, we are able in our experiments to systematically control the magnitude of $d_3$ and to investigate the soliton dynamics both when driving in the anomalous and normal dispersion regimes~\cite{li_experimental_2020}.

\begin{figure*}[!t]
	\centering
	\includegraphics[scale=1]{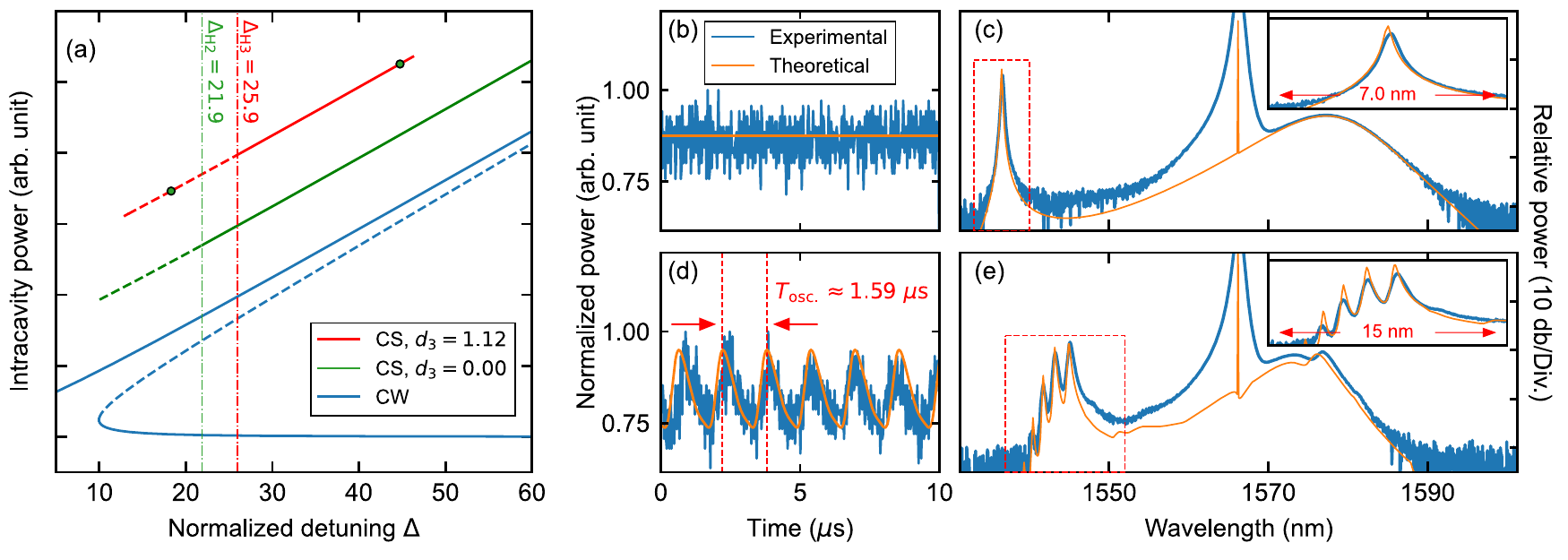}
	\caption{\small(a) Red curve shows CS bifurcation curve for X = 150 and $d_3 = 1.12$; green curve shows corresponding results in the absence of higher-order dispersion [$\beta_3 = \beta_4 = 0$] and stimulated Raman scattering [$f_\mathrm{R} = 0$]. Vertical dashed-dotted lines labelled as $\Delta_{\mathrm{H}2}$ and $\Delta_{\mathrm{H}3}$ indicate the detunings at which breathing solitons emerge in the absence and presence of higher-order terms, respectively. Blue curves show CW solutions for X = 150. Solid (dashed) curves correspond to stable (unstable) solutions. (b) and (c) blue (orange) curves show experimentally measured (numerically simulated) CS energy per round-trip and time-averaged spectrum at $\Delta = 44.5$, respectively. (d) and (e) are as in (b) and (c), respectively, but with $\Delta = 18$. The positions of both measurements along the CS bifurcation curves are indicated in (a) by green circles. Inserts in (c) and (e) show the dispersive wave spectral components in more detail.}\label{single_sol}
\end{figure*}

\section{Results}
\subsection{Anomalous-dispersion driving}
We first consider experiments that probe the CSs' instability dynamics under near-zero-dispersion conditions when driving in the anomalous dispersion regime. To this end, we set the pump power to $0.97$~W such that $X = 150$, and we set the driving wavelength to $1566.2$~nm (i.e., 1.7~nm away from the ZDW), where dispersion is anomalous and \mbox{$d_3 = 1.12$}. In Fig.~\ref{single_sol}(a), we show a theoretically predicted CS bifurcation curve [red curve] obtained by finding the steady-state solutions of Eq.~\eqref{LLE} using a Newton-Raphson method with our experimental parameters [see also Figure caption]. Also shown in green is the corresponding curve in the absence of higher-order terms [i.e., $\beta_k = 0$ for $k>2$  and $f_\mathrm{R} = 0$], as well as the intensity levels of the homogeneous CW state of the system. (Note that the stability of all the solutions was deduced via a linear stability analysis~\cite{kuznecov_elements_2010}.) We see that, both in the presence and absence of higher-order dispersion terms, the CSs are stable (solid curves) for large detunings, but then become unstable (dashed curves) through a Hopf bifurcation at detuning $\Delta_{\mathrm{H}3}$ and $\Delta_{\mathrm{H}2}$, respectively.

%As can be seen, both in the presence and absence of higher-order terms, the CSs are stable (solid curves) for large detunings, but then become unstable (dashed curves) through a Hopf bifurcation at detunings $\Delta_{\mathrm{H}3}$ and $\Delta_{\mathrm{H}2}$, respectively.

\begin{figure}[!h]
	\centering
	\includegraphics[scale=1]{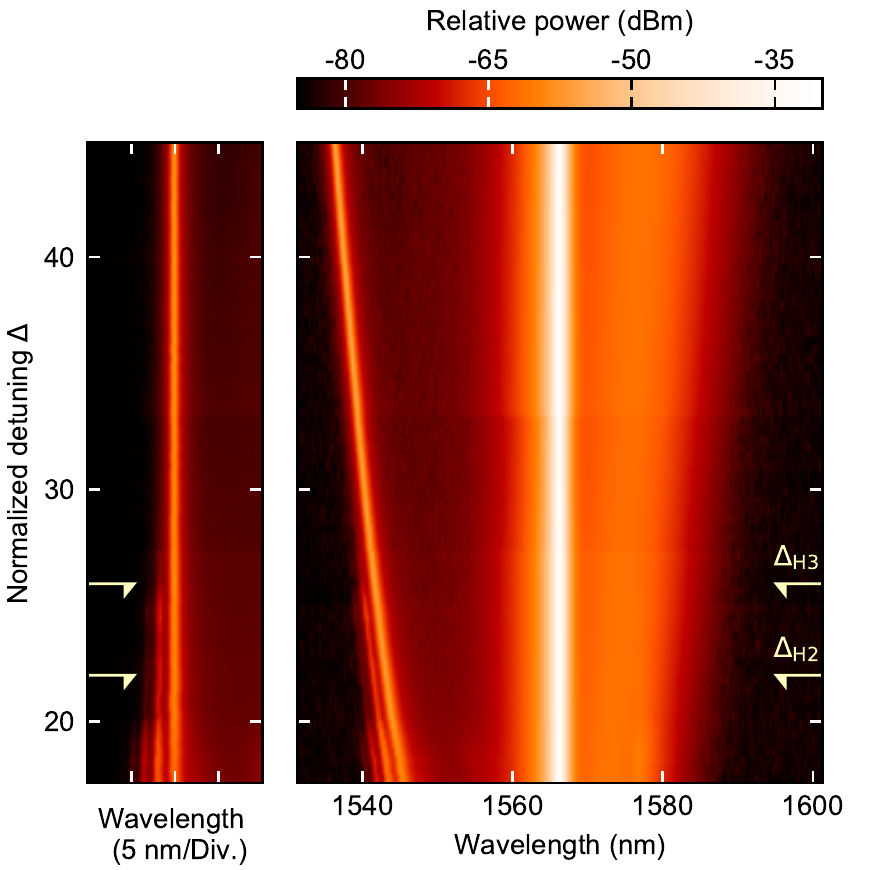}
	\caption{\small The right panel shows a pseudo-colour plot of experimentally measured spectra at various detuning values, obtained through vertically concatenating 72 individual spectral measurements with parameters as in Fig.~\ref{single_sol}. The left panel shows a corresponding plot of high resolution spectral measurements around the dispersive wave spectral profile; note that the latter measurements are centred at the wavelength of the strongest dispersive radiation peak and encompass a spectral window of 20~nm. In both panels, the predicted positions of the CS Hopf bifurcation points in the presence and absence of higher-order terms, $\Delta_{\mathrm{H}3}$ and $\Delta_{\mathrm{H}2}$ [same values shown in Fig.~\ref{single_sol}(a)], are indicated by the yellow arrows, respectively.}\label{1566_scan}
\end{figure}

In our experiments, we stabilize the detuning at different values and examine the soliton characteristics. Blue curves in Figs.~\ref{single_sol}(b)--(e) show the experimentally measured CS energy per roundtrip as well as time-averaged optical spectra when the detuning is stabilised at \mbox{$\Delta = 44.5$} [Figs.~\ref{single_sol}(b, c)] and \mbox{$\Delta = 18$} [Figs.~\ref{single_sol}(d, e)]. In agreement with the bifurcation curve shown in Fig.~\ref{single_sol}(a), the CSs at $\Delta = 44.5$ are dynamically stable, exhibiting a typical spectrum with a strong monochromatic dispersive wave peak at about 1537~nm. In stark contrast, at $\Delta = 18$, the solitons exhibit regular oscillations with a period of about $T_\mathrm{osc} = 1.59~\mu\mathrm{s}$ (or equivalently 70 roundtrips), with the dispersive wave peak exhibiting a manifestly polychromatic character. The latter feature was recently predicted by Melchert et al.~\cite{melchert_multi-frequency_2020} and can be explained as follows. Efficient dispersive wave emission requires that the CS spectrally overlaps with the phase-matched dispersive wave wavelength\cite{webb_efficiency_2014}; when the (spectral width of the) CS oscillates, the efficiency of dispersive wave emission is accordingly modulated. Compounded by the fact that the emitted dispersive waves walk off from the soliton due to dispersion, this results in the generation of a quasi-periodic train of phase coherent dispersive waves with decaying amplitude. The temporal spacing between two consecutive dispersive waves, $\Delta\tau$, is equal to the dispersive walk-off during one soliton oscillation period $T_\mathrm{osc.}$, i.e.,
\begin{equation}
\Delta\tau = L\frac{T_\mathrm{osc}}{t_\mathrm{R}}\left[\beta_\mathrm{2,s}\Omega + \frac{\beta_\mathrm{3,s}}{2}\Omega^2 + \frac{\beta_\mathrm{4,s}}{6}\Omega^3  \right],
\end{equation}
 where $\Omega$ is the angular frequency shift of the dispersive wave from the soliton and the dispersion coefficients $\beta_{k,\mathrm{s}}$ are evaluated at the center of the soliton spectrum. For our parameters (estimating the soliton and the dispersive wave to be centred at 1573~nm 1543~nm, respectively) $\Delta\tau = \Delta \tau ^{-1} \approx 4.9~\mathrm{ps}$, yielding a corresponding spectral modulation frequency $\Delta f \approx 204~\mathrm{GHz}$, which is in reasonable agreement with 198~GHz extracted from the measured spectrum (average frequency spacing between the peaks that are not equidistant due to dispersion). To the best of our knowledge, the results shown in Figs.~\ref{single_sol}(d) and (e) represent the first experimental confirmation of polychromatic dispersive wave emission by oscillating near-zero-dispersion CSs.

The experimental observations in Figs.~\ref{single_sol}(b)--(e) are in good agreement with numerical simulations [see orange curves in Fig.~\ref{single_sol}] and predictions of the theoretical bifurcation curve [Fig.~\ref{single_sol}(a)]. Somewhat surprisingly, the bifurcation curve predicts that, in the presence of higher-order terms, the Hopf bifurcation that marks the transition between stable and unstable CSs occurs at a larger detuning than in the pure LLE model, i.e., $\Delta_{\mathrm{H}3} > \Delta_{\mathrm{H}2}$. Arising predominantly due to third-order dispersion (we again emphasize that fourth-order dispersion and Raman scattering play a minor role only), this observation seemingly contradicts the notion that third-order dispersion stabilises CSs~\cite{parra-rivas_third-order_2014}. To experimentally test this prediction, we measured the soliton characteristics as a function of detuning over the entire range of the soliton existence. More specifically, we excite and stabilise a soliton at a given detuning, and then adiabatically adjust the detuning stabilisation point in small steps while recording the soliton spectral and temporal profiles. (Of note: occasionally the soliton dies due to environmental perturbations during the process, in which case we re-excite the soliton and resume the measurement.) Figure~\ref{1566_scan} shows the measured soliton spectral evolution with $X = 150$ [i.e. parameters as in Fig.~\ref{single_sol}(a)], and reveals that the dispersive wave component transforms from poly- to monochromatic at a detuning $\Delta\approx 26$. This transformation is indicative of the transition between unstable and stable CSs, and the detuning at which it occurs is in good agreement with the predicted Hopf bifurcation point $\Delta_{\mathrm{H}3} = 25.9$. (Note that stability is inferred from the presence of polychromatic dispersive waves in the spectrum, which provides a better signal-to-noise ratio than temporal measurements.) The observed transition point is noticeably larger (the uncertainty in our measurement of $\Delta$ is about unity) than the value of $\Delta_{\mathrm{H}2} = 21.9$ predicted in the pure LLE model, thus confirming that, for the parameters considered in Fig.~\ref{single_sol} and \ref{1566_scan}, third-order dispersion does \emph{not} stabilize the CS.

\begin{figure}[!h]
	\centering
	\includegraphics[scale=1]{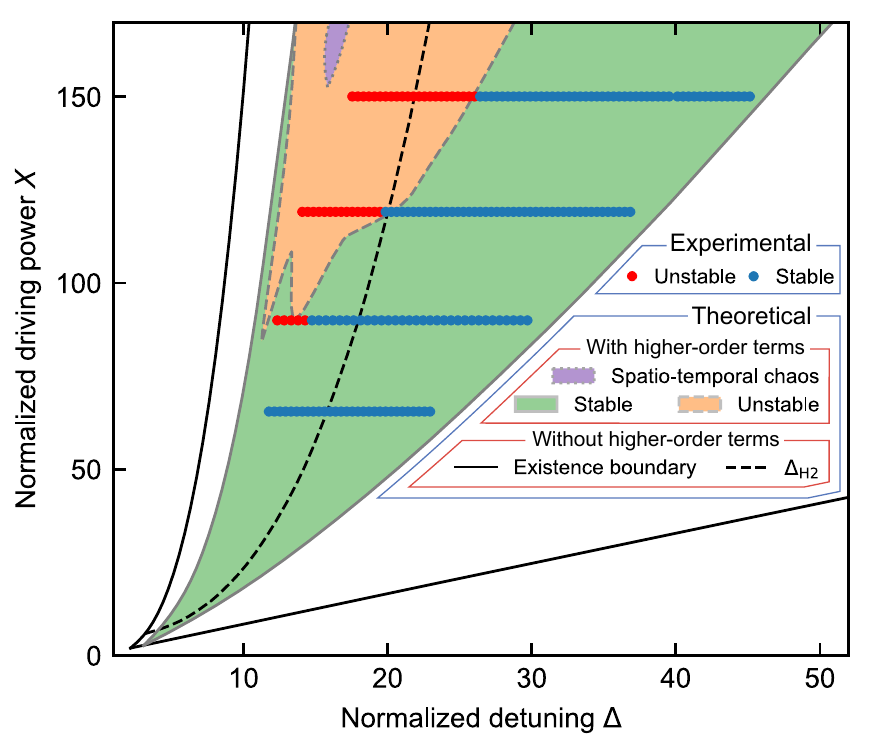}
	\caption{\small Bifurcation characteristics of CSs with parameters as in Fig.~\ref{single_sol} over a range of normalized detunings and driving powers; solutions within the green, orange and purple patches are stable, oscillatory and spatio-temporally chaotic, respectively. Black solid lines delineate the existence boundaries of CSs in the absence of higher-order terms (pure LLE model) whilst the black dashed line shows the corresponding Hopf bifucation point $\Delta_{\mathrm{H}2}$. Blue and red dots correspond to experimental measurements of stable and unstable CSs, respectively.}\label{1566_roe}
\end{figure}

To gain more insight and explain the apparent contradiction between our findings and results reported in Ref.~\cite{parra-rivas_third-order_2014}, we computed the soliton's bifurcation characteristics for a range of driving powers and detunings. The results are shown in Fig.~\ref{1566_roe} for $d_3 = 1.12$, where various coloured regions indicate parameters space over which solitons exhibit distinct dynamical behaviours, superimposed with the bifurcation points predicted in the absence of higher-order effects. We have carefully verified that neglecting fourth-order dispersion and Raman scattering yields qualitatively similar existence and stability boundaries, thus allowing us to draw two major conclusions with regards to the effect of third-order dispersion.  First, third-order dispersion reduces the range of detunings over which the CS exists -- especially on the large detuning side. Second, third-order dispersion induces a non-trivial change in solitons' stability boundaries. For low driving power level ($X < 80$), the solitons are stable at all detunings -- even when instabilities would otherwise manifest themselves in the pure LLE model. Thus, for low driving powers, third-order dispersion stabilises the solitons, as predicted in Ref.~\cite{parra-rivas_third-order_2014}. As the driving power increases, we find however that the solitons' immunity against instabilities vanishes, and we observe a cross-over where solitons are unstable over a larger range of detunings in the presence of third-order dispersion than in its absence. The results in Fig.~\ref{1566_roe} suggest that the stabilising effect of third-order dispersion (as reported in Ref.~\cite{parra-rivas_third-order_2014}) only holds for low driving power levels, whilst for larger power levels, the instability range is, in fact, extended [congruent with results reported in Fig.~\ref{single_sol} and Fig.~\ref{1566_scan}].

To more comprehensively test the bifurcation predictions, we performed experiments as in Fig.~\ref{1566_scan} for a range of driving power levels. The results are shown in Fig.~\ref{1566_roe} with blue and red solids circles representing stable and oscillatory solitons, respectively. Relating to the instability transition, the experimental results are in good agreement with the theoretical predictions. There is only one discrepancy: a theoretically predicted small island of stability around $X = 100$ is not observed in our experiment. We suspect this is due to experimental imperfections that prevent access to the small island predicted. (Of note: we have observed the corresponding island in experiments performed for larger values of $d_3$ where the island extends over a larger range of detuning.) Relating to the range of existence, our experiments show good qualitative agreement with numerical predictions, but quantitative discrepancies arise at the low detuning side. We believe this may be due to inhomogeneities in the driving pulse that locally trigger modulation instability that then pervades the entire intracavity field. Nonetheless, our experiments confirm the salient theoretical predictions: third-order dispersion limits the range of soliton existence and switches from stabilizing to destabilizing for large driving power levels.

\subsection{Normal-dispersion driving}

The results reported above pertain to the (common) situation where the resonator exhibits anomalous dispersion at the driving wavelength. However, recent studies have shown that, due to higher-order dispersion, the existence of bright solitons can extend into the regime of normal-dispersion-driving~\cite{milian_soliton_2014,li_experimental_2020}. As a matter of fact, a number of different bright structures can exist when driving close to the ZDW in the normal dispersion regime, distinguished by a different number of prominent peaks in their temporal profile~\cite{parra-rivas_coexistence_2017}. Here we are interested in the ``single-peak'' structures that are the normal-dispersion counterparts of the standard CSs that exist in the anomalous dispersion regime~\cite{li_experimental_2020}; we will henceforth refer to such structures simply as CSs.

\begin{figure}[!h]
	\centering
	\includegraphics[scale=1]{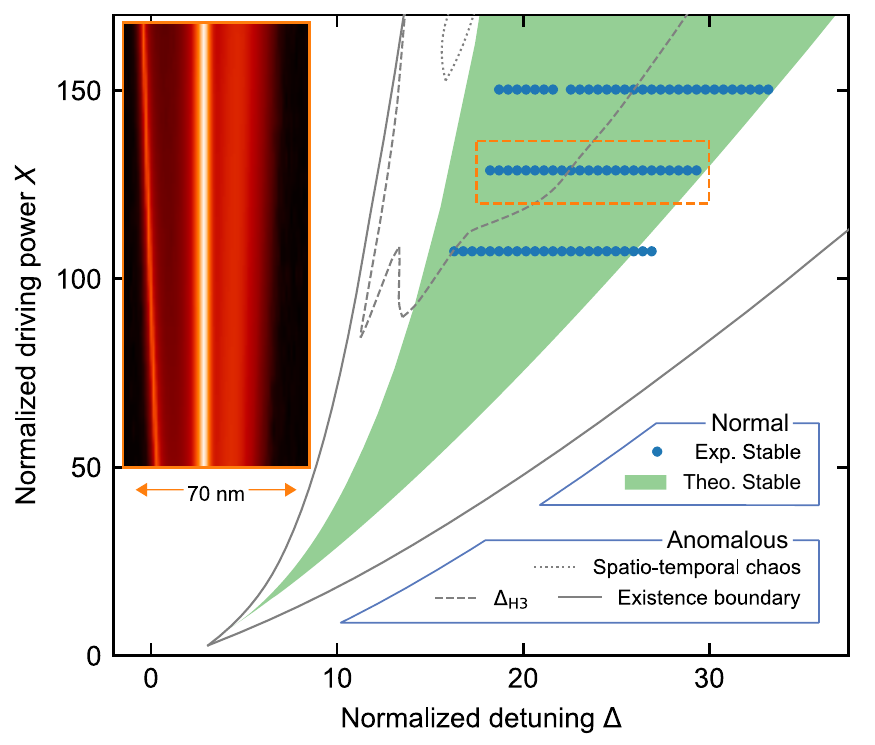}
	\caption{\small Bifurcation characteristic of single-peak CSs with normal-dispersion driving and $d_3 = 1.07$ over a range of normalized detunings and driving powers. The green patch indicates the parameter range within which normal-dispersion CS exist and are predicted to be stable; note that there are no instabilities for the considered parameters. For comparison, the solid grey curves delineate the existence boundaries of anomalous-dispersion CSs with $d_3 = 1.12$, whilst dashed and dotted grey curves indicate the corresponding instability boundaries (data extracted from Fig.~\ref{1566_roe}). Blue dots correspond to experimental measurements of stable CSs (no unstable CSs were observed for the considered parameters). The insert shows a pseudo-colour plot of spectral measurements at $X = 120$ and for various detunings (corresponding to data points enclosed by the dashed orange rectangle). Note that, due to experimental error, a data point near $X = 150$ and $\Delta = 20$ is missing. However, by observing the stability state of its neighbours, it is reasonable to assume the CS at this position will be stable.}\label{1562_roe}
\end{figure}

To explore the existence and stability of CSs with normal-dispersion driving, we adjust the driving wavelength to 1562.8~nm (i.e., 1.6~nm away from the ZDW) where the group-velocity dispersion is normal and $d_3 = 1.07$. We then repeated the theoretical and experimental analyses reported in Fig.~\ref{1566_roe} and summarise the results in Fig.~\ref{1562_roe}. Also shown for comparison are the theoretical boundaries in the presence of higher-order terms as extracted from Fig.~\ref{1566_roe}.

Several conclusion can be drawn. First, the solitons' range of existence is further reduced from the case of anomalous-dispersion driving. Second, in contrast to CSs in the anomalous dispersion regime, the solitons are stable for all the detunings and driving powers considered. (Of note: simulations predict that instabilities can arise if one consider even larger driving levels.) Third, our experiments are again in good qualitative agreement with the theoretical results, showing no signs of instability and reproducing the theoretically predicted upper-detuning limit of soliton existence. As for the anomalous dispersion results (Fig.~\ref{1566_roe}), discrepancies arise especially at the low-detuning limit. It is interesting to note that normal-dispersion driving appears to permit full immunity against instabilities over a larger range of driving powers than anomalous-dispersion driving. However, it is important to stress that, at each driving power, the latter regime is still associated with a larger overall range of detunings where the solitons are stable thanks to the larger range of existence of CSs in the anomalous dispersion regime.

\subsection{Zero-dispersion driving}

The results reported in Figs.~\ref{1566_roe} and \ref{1562_roe} show that, despite similar relative third-order dispersion parameters $d_3 \approx 1$, the existence and stability of CSs differ markedly between anomalous and normal dispersion driving. On the one hand, this is within expectation since CSs are commonly associated with anomalous dispersion driving only (and do no exist at all with normal dispersion driving in the limit of $d_3  \rightarrow 0$). But on the other hand, one may expect that in the limit $d_3 \rightarrow \infty$, which occurs as the pump approaches the ZDW of the resonator, the two driving scenarios must converge to one; indeed, in this case, the second-order dispersion can be considered negligible and the driving scenarios physically merge. 

\begin{figure}[t]
	\centering
	\includegraphics[scale=1]{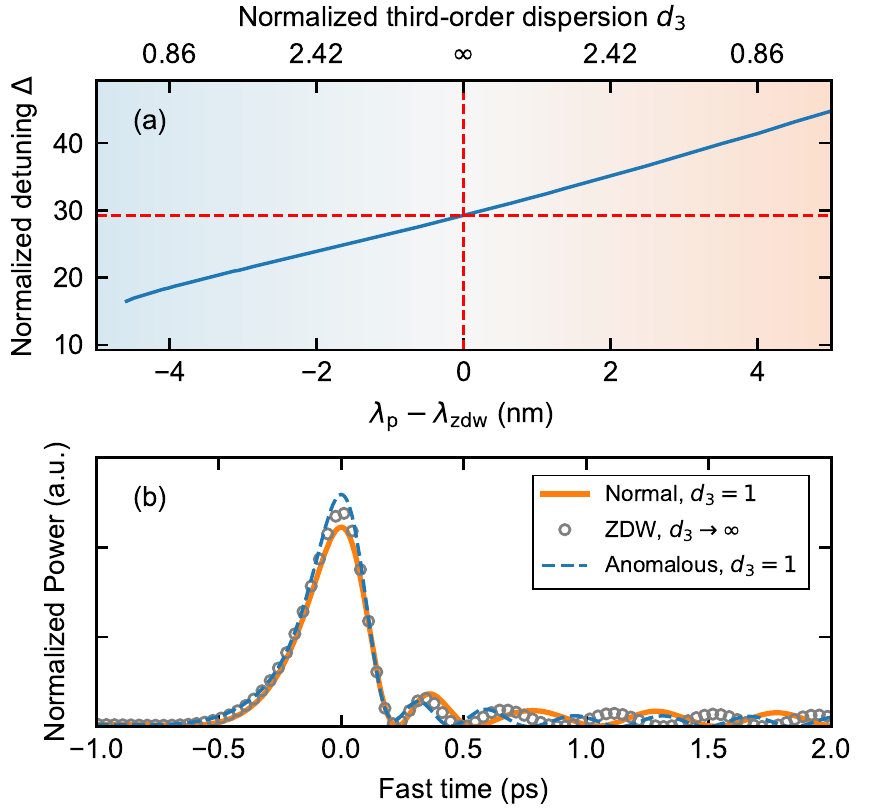}
	\caption{(a) Upper-detuning-limit of soliton range of existence as a function of deviation from the ZDW, $\lambda_\mathrm{p} - \lambda_\mathrm{zdw}$, where $\lambda_\mathrm{p}$ and $\lambda_\mathrm{zdw}$ are the driving wavelength and ZDW, respectively. The dashed red line indicates the detuning-limit of the zero-dispersion soliton that exists when $\beta_2 = 0$ (or equivalently $d_3 = \infty$). (b) Temporal profiles of simulated CSs at $\Delta = 20$ for normal-dispersion driving (orange solid curve), anomalous-dispersion driving (blue dashed curve), and driving exactly at the ZDW (grey circles)}\label{convergence}
\end{figure}

To elucidate whether the soliton dynamics evolve smoothly as the driving wavelength is tuned across the ZDW (or whether a separatrix exists that divides the two regions), we theoretically calculated the soliton solutions of Eq.~(\ref{LLE}) over a wide range of driving wavelengths on either side of the ZDW for a constant $X = 100$. To compare the soliton properties, we plot in Fig.~\ref{convergence}(a) the normalized cavity detuning where the solitons cease to exist (the upper limit of existence) as a function of pump wavelength (and hence relative third-order dispersion). As can be seen, the upper-detuning-limit varies smoothly across the ZDW, monotonically decreasing as the pump wavelength is moved from anomalous to normal dispersion regime, until the soliton eventually ceases to exist in an abrupt fashion when the pump is tuned sufficiently deep into the normal dispersion regime. This result corroborates the observation made in Ref.~\cite{li_experimental_2020} that there is no qualitative change in the characteristic of near-zero-dispersion CSs across the ZDW. 

Before closing, we finally note that when the pump wavelength is close to the ZDW, the dynamics and characteristics of the CSs can of course be well-predicted by Eq.~(\ref{LLE}) with second-order dispersion identically set to zero, i.e. $\beta_2 = 0$. The dashed red lines in Fig.~\ref{convergence}(a) highlight the upper-detuning-limit of existence of solitons that exist when $\beta_2 = 0$, whilst Fig.~\ref{convergence}(b) shows the corresponding soliton temporal profile compared with selected profiles for normal and anomalous dispersion driving with $\Delta = 20$ and $d_3 = 1$. A thorough analysis of the characteristics and dynamics of such single-peak, zero-dispersion CS is beyond the scope of our present work, yet we envisage that such future analysis can provide valuable insights and facilitate the design of ultra-broadband microresonator frequency combs and sources of ultra-short cavity solitons.    

\section{Conclusions}

In conclusion, we have experimentally studied the impact of third-order dispersion on the existence and stability boundaries of temporal Kerr cavity solitons. This has been achieved by using an experimental configuration that allows operation (arbitrarily) close to the ZDW of the resonator, thus enabling systematic control over the relative impact of third-order dispersion. Our results include the first observations of polychromatic dispersive waves emitted by breathing near-zero-dispersion CSs, as predicted by earlier theories~\cite{melchert_multi-frequency_2020}. Moreover, we have shown that, for low driving powers, third-order dispersion can stabilize CSs (as predicted by earlier theories~\cite{parra-rivas_third-order_2014}), but that for high-power levels, the effect of third-order dispersion switches from stabilizing to destablizing, expanding the range of parameters over which solitons are unstable. In addition to experiments performed under conditions of anomalous-dispersion driving, we have also considered solitons that manifest themselves with the driving field in the normal dispersion regime, and examined how the two pumping conditions merge as the driving wavelength approaches the ZDW. Our experiments are in good agreement with numerical simulations of the generalized Lugiato-Lefever equations, with some discrepancies likely stemming from experimental imperfections. Our work provides convincing experimental confirmation of earlier theoretical predictions~\cite{melchert_multi-frequency_2020, parra-rivas_third-order_2014} and provides new insights on CSs with (near-)zero-dispersion driving.

\section{Acknowledgements}
The authors acknowledge financial support from the Marsden Fund and the Rutherford Discovery Fellowships of the Royal Society of New Zealand.

%\bibliographystyle{bibstyle2nonotes}
%\bibliography{citation,breather}

\end{document}